\documentstyle[preprint,aps,prl]{revtex}

\begin{document}

\draft

\title{
Quasi-Langevin method for shot noise calculation in single-electron 
tunneling}

\author{Alexander N. Korotkov}
\address{
Nuclear Physics Institute, Moscow State University, 
Moscow 119899, Russia, \\
NEC Fundamental Research Laboratories, Tsukuba, Ibaraki 305, Japan, \\
and Department of Physics, State University of New York,
Stony Brook, NY 11794-3800 }

\date{\today}

\maketitle

\begin{abstract}
        It is shown that quasi-Langevin method 
can be used for the calculation of the 
shot noise in correlated single-electron tunneling.
        We generalize the existing Fokker-Plank-type approach
and show its equivalence to quasi-Langevin approach.
        The advantage of the quasi-Langevin method is a natural
possibility to describe simultaneously the high (``quantum'') 
frequency range.

\end{abstract}

\pacs{73.40.Gk; 72.70.+m}

\narrowtext

\vspace{1ex}

        Correlated single-electron tunneling \cite{Av-Likh} remains
an attractive topic during last decade. Because in the systems of
small-capacitance high-resistance tunnel junctions electrons tunnel
almost as the classical particles, the most of experiments are well
explained within the framework of ``orthodox'' theory \cite{Av-Likh}
of single-electronics which is based on the classical master equation.

        The theory of the shot noise in single-electron 
transistor \cite{Likh-87} developed in Refs.\  
\cite{SQUID,Thesis} is also based on the master equation. 
The shot noise appears due to the randomness of the tunneling events. 
Despite the classical description of the system, the current 
in this theory should be treated as a kind
of an operator because the current is caused by tunneling events
which change the charge state of the system. 
This theory has been studied in detail in a number of subsequent 
publications - see, e.g., Refs.\ 
\cite{Hershfield,Chen,Galperin,Krech,Galperin-SC,Kor-sc,Zorin}.
The first experimental confirmation of the theory has been obtained 
\cite{Birk} in 1995, and other experiments are definitely coming
because several groups has succeeded in the fabrication of the
single-electron transistor operating at relatively high 
frequencies (hundreds of kHz) at which 1/f noise should 
become less than the shot noise.

        Among the most noticeable developments of the shot noise
theory in single-electronics after 1991, let us mention
the following. It has been shown \cite{Galperin-SC} that the noise 
due to the processes of Andreev reflection
which transfer two electrons per tunneling event
can be described  by the same formalism with $e$ substituted by $2e$.
The shot noise in single-electron transistor with large
level discreteness has been studied \cite{Averin-noise}.
The shot noise theory has been applied to single-electron systems 
other than single-electron transistor \cite{Kor-Bl,Kor-oscil}. 
Besides the noise in the ``orthodox'' frequency range
$\omega \sim W/e^2 R$ (where $W$ is a typical energy and 
$R$ is a typical resistance), the noise in the ``quantum'' frequency 
range $\omega \sim W/\hbar$ has been 
studied and the matching of the high and low frequency limits of
two ranges has been proven    
\cite {Kor-Bl}; however, the approach unifying both  
frequency ranges in one formalism has not been found. 

        The existing formalism of the shot noise in single-electron 
tunneling is of the Fokker-Plank type and is based on the deterministic
master equation. However, generally more popular method in the study
of the noise is the Langevin approach 
in which the random term is introduced into the evolution
equation. Three decades ago the Langevin approach was applied 
\cite{Kogan} to study the fluctuations in the nonequilibrium electron
gas. In the present letter we show that a similar method can be used
for the  single-electron  tunneling. (We call it quasi-Langevin
because the regular Langevin method cannot be applied directly.)
  It  is  equivalent  to   existing 
formalism in the ``orthodox'' frequency range,  
but it naturally allows also the calculations
in the ``quantum'' frequency range.

        Let us start with the generalization of the existing 
Fokker-Plank type method to an arbitrary system consisting of 
voltage sources, 
capacitances and tunnel junctions with sufficiently large resistances 
($R_j \gg R_K=h/e^2\simeq 26\mbox{k}\Omega$). 
For simplicity we do not consider Ohmic
resistances. In this case the dynamics is governed by the matrix master
equation
        \begin{equation}
{\dot {\bf \sigma}} = {\bf \Gamma} {\bf \sigma}
        \label{mastereq}\end{equation}
        where the vector $\sigma_n$ is the probability to find the 
system 
in the charge state $n \equiv \{n_1,...n_L\}$ (which is characterized 
by the numbers $n_i$ of excess electrons in each of $L$ internal nodes 
of the system) and
        \begin{equation}
\Gamma_{mn} = \Gamma_{m\leftarrow n} - \delta_{mn} \sum_k 
\Gamma_{k\leftarrow n}, \,\, 
\Gamma_{m\leftarrow n} = \sum_j \Gamma^j_{m\leftarrow n} ,
        \label{Gamma}\end{equation}
        where $\Gamma_{m\leftarrow n}^j$ are the corresponding tunneling 
rates and the summation over the junction number $j$ is necessary
when an electron can tunnel to (from)   
an internal node from (to) different external electrodes.

        To find the mutual spectral density for two processes
$X(t)$ and $Y(t)$ we can calculate first the correlation function
$K_{XY}(\tau )=\langle X(t+\tau )Y(t)\rangle -\langle X \rangle
\langle Y\rangle$ (brackets denote the averaging
over time) and then take the Fourier transform
        $S_{XY}(\omega )=2\int_{-\infty}^{+\infty} K_{XY} (\tau ) \exp 
(i \omega \tau) \, d\tau$ .
        If both $X$ and $Y$ are functions of the charge state $n$ 
(for example, potential of a node) then the correlation function is
given by the simple expression 
        \begin{eqnarray}
K_{XY}(\tau ) =  \theta (\tau ) \sum_{m,n} X(m) \, 
\sigma(\tau , m|n) \, Y(n)\, \sigma_{n}^{st} 
        \nonumber \\
+\, \theta (-\tau )\sum_{m,n} Y(m) \, \sigma (m,-\tau |n) \, X(n)\,
 \sigma_n^{st}-\langle X\rangle \langle Y\rangle \, , 
        \label{K-simple}\end{eqnarray}
        where $\sigma(\tau ,m|n)$ is the retarded Green's function
of Eq.\ (\ref{mastereq}) being the probability to find the system
in the state $m$ at $t=\tau >0$ if at $t=0$ it was in the state $n$,
$\langle X \rangle=\sum_n X(n)\sigma_n^{st}$, 
and $\sigma_n^{st}$ is the stationary distribution, ${\bf \Gamma} 
{\bf \sigma}^{st}=0, \,\, \sum_n\sigma_n^{st}=1$. (Notice that $X$ and $Y$
are classical variables, and their commutator is zero.)

        However, if $X$ and/or $Y$ represent the current through a 
tunnel junction or in an external lead, the Eq.\ (\ref{K-simple})
should be modified. For example, if $X(t)$ is the current contribution
corresponding
to tunneling events $\Gamma_{m\leftarrow n}^j$ while $Y(t)$ 
corresponds to $\Gamma_{m'\leftarrow n'}^{j'}$, then (similar to 
Refs.\ \cite{SQUID,Thesis})      
        \begin{eqnarray}
K_{XY}(\tau )/{\tilde e}^{j}_{\pm}{\tilde e}^{j'}_{\pm} =  
\theta (\tau ) \, \Gamma_{m\leftarrow n}^j \,
\sigma(\tau ,n|m') \, \Gamma_{m'\leftarrow n'}^{j'}\sigma_{n'}^{st} 
        \nonumber \\
+\, \theta (-\tau ) \, \Gamma_{m'\leftarrow n'}^{j'} \sigma (n',-\tau |m) 
\, \Gamma_{m\leftarrow n}^j \, \sigma_n^{st}
        \nonumber \\
- \, \Gamma_{m\leftarrow n}^j \sigma_n^{st} \, \Gamma_{m'\leftarrow n'}^{j'}
\sigma_{n'}^{st}  
+ \delta_{mm'}\delta_{nn'}\delta_{jj'} \delta (\tau) \,
\Gamma_{m\leftarrow n}^j  \sigma_n^{st} \, . 
        \label{K-current}\end{eqnarray}
Here the last term is responsible for the high-frequency limit.
The effective charges ${\tilde e}^{j}_{\pm}$ and ${\tilde e}^{j'}_{\pm}$
are determined by the direction of electron tunneling, 
${\tilde e}^j_+=-{\tilde e}^j_-$, and by the circuit capacitances
\cite{SQUID,Thesis} (so that ${\tilde e}^{j}=e$ only if the current 
through junction $j$ is measured).  
        Any current--current correlation function can be written as a
sum of $K_{XY} (\tau)$ given by Eq.\ (\ref{K-current}) 
over all possible transitions between
charge states (such a sum is a counterpart of Eq.\ (\ref{K-simple})
in which the sum is written explicitly). The processes of cotunneling
\cite{Av-Odin} in the simple master equation approximation 
\cite{Fonseca} and fluctuations due to Andreev reflection 
\cite{Galperin-SC} can be accommodated in this technique using
the appropriate values of ${\tilde e}^j$ (for cotunneling the index
$j$ should obviously represent the set of junction numbers).

        For the correlation functions when $X$ is a current and $Y$
is a function of the charge state (or vice versa), the recipe 
is the ``combination'' of Eqs.\ (\ref{K-simple}) and (\ref{K-current})
while the term proportional to $\delta (\tau )$ is absent.

        The expressions for spectral densities directly follow from 
Eqs.\ (\ref{K-simple}) and (\ref{K-current}) because the  Fourier 
transformation affects only the evolution operator 
$\sigma(\tau ,m|n)$, and the corresponding Green's function in the 
frequency representation is simply obtained from Eq.\ (\ref{mastereq})
        \begin{equation}
\sigma (\omega ,m|n)= \left[ (-i\omega {\bf 1} -{\bf \Gamma})^{-1} 
\right]_{mn}  ,
        \label{sigma(w)}\end{equation}
where ${\bf 1}$ is the unity matrix. 
        For example, Eq.\ (\ref{K-current}) leads to the following
spectral density
        \begin{eqnarray}
S_{XY} (\omega )/{\tilde e}^{j}_{\pm}{\tilde e}^{j'}_{\pm} = 
2 \, \Gamma^j_{m\leftarrow n} \left[
(-i\omega {\bf 1} -{\bf \Gamma})^{-1} \right]_{nm'} 
\Gamma^{j'}_{m'\leftarrow n'} \sigma_{n'}^{st} 
        \nonumber \\
+ \,  2 \, \Gamma^{j'}_{m'\leftarrow n'}
 \left[ (i\omega {\bf 1} -{\bf \Gamma} )^{-1}  
\right]_{n'm} \Gamma^{j}_{m\leftarrow n}
\sigma_{n}^{st} 
        \nonumber \\
+\, 2 \, \delta_{nn'}\delta_{mm'}\delta_{jj'} \,
\Gamma^{j}_{m\leftarrow n} \, \sigma_{n}^{st} .
        \label{SII}\end{eqnarray}

        This method \cite{SQUID,Thesis} allows to calculate all 
spectral densities within
the framework of ``orthodox'' theory, and at least for the single-electron
transistor the numerical procedure is rather trivial because matrix
${\bf\Gamma}$ is three-diagonal and the calculation of Eq.\ (\ref{sigma(w)})
is straitforward \cite{complicat}. There are some additional 
simplifications for $\omega \rightarrow 0$, however, in this paper 
we concentrate on the finite frequencies only.
	Notice that in the special case when there are only two charge 
states involved, the theory is equivalent to the noise calculations 
in resonant tunneling diode \cite{Chen-Ting}.
 
        Now let us develop the Langevin-type approach. It is obvious that
for the discrete random dynamics of the charge states of a  
single-electron circuit 
it is impossible to introduce the random term in Eq.\ (\ref{mastereq}) 
in a reasonable way.
However, let us imagine the ensemble of $M$ ($M\gg 1 $) independent  
similar circuits, and let us average all magnitudes over this ensemble.
Then the average (over time) currents and voltages will not change
(due to ergodicity), but the spectral densities of fluctuations
(second order magnitudes) will decrease $M$ times. Hence, to calculate 
spectral densities of the initial system, we can take the leading 
($\sim M^{-1}$) order of the spectral density of magnitudes 
averaged over the ensemble, in the 
thermodynamic limit $M\rightarrow \infty$.
In contrast to the single system, the dynamics of the 
ensemble at $M\rightarrow \infty$ can be easily described by
the Langevin approach because there are no longer step-like 
dependencies on time. 

        Now $M \sigma_n(t)$ which is the number of systems 
in the ensemble having the charge 
state $n$, can be considered as non-integer (because of $M\gg 1$)
and its random dynamics can be
described with the help of fictitious random term $\xi (t)$.
        In the ``orthodox'' framework (when we consider electron 
jumps as instantaneous events), $\xi (t)$ is $\delta$-correlated 
(white), and its amplitude is defined by the Poissonian nature 
of the tunneling process. The recipe is the following \cite{Kogan}:
for each average flux $M \Gamma_{m\leftarrow n}^j \sigma^{st}_n$ 
(number of transitions per second) in the space of charge states,
we should add in the master equation  the random 
flux $\xi_{m\leftarrow n}^j (t)$ with the corresponding ``seed'' spectral 
density given by usual Schottky formula 
        \begin{eqnarray}
{\dot \sigma}_m (t) =\sum_n  \Gamma_{mn} \sigma_n (t) + \xi_m(t),
        \label{mas-eq2} \\
\xi_m(t) =
\sum_{n,j} \xi_{m\leftarrow n}^j (t) - \xi_{n\leftarrow m}^j(t),
        \label{xim} \\
S_{\xi_{m\leftarrow n}^j \xi_{m'\leftarrow n'}^{j'}} (\omega) = 
2 M^{-1} \delta_{mm'}\delta_{nn'}\delta_{jj'}  \Gamma_{m\leftarrow n}^j 
\sigma_n^{st}.
        \label{Sxi}\end{eqnarray}
 Notice that when there are
fluxes in opposite directions ($m\leftarrow n$ and
$n\leftarrow m$), we should apply $\xi (t)$ for each direction, so
that the random flux does not vanish even if the net 
average flux is zero. 

        Because of the linearity of Eqs.\ (\ref{mas-eq2})--(\ref{Sxi})
the final spectral densities of the averaged (over $M$) 
magnitudes are obviously proportional to $1/M$. Hence, rescaling to the 
single system can be done formally assuming $M=1$ in 
Eqs.\ (\ref{mas-eq2})--(\ref{Sxi}). So, instead of keeping $M$ and
rescaling at the final stage, we will use $M=1$ in all equation
below. We call this simple trick of considering first the large 
ensemble, writing the Langevin equation for it, and then returning
to a single system, a quasi-Langevin approach.

        Using the standard procedure we find the Fourier transform
        \begin{equation}
\sigma _m (\omega) =\left [ (-i\omega {\bf 1} -{\bf \Gamma} )
^{-1} \right]_{mn} \xi_n (\omega ). 
        \label{sigma}\end{equation}
Then for the occupation--occupation spectral density
        \begin{eqnarray}
S_{\sigma_m \sigma_n}=\sum_{m'n'} 
\left[ (-i\omega {\bf 1} -{\bf \Gamma})^{-1}\right]_{mm'} 
        \nonumber \\
\times \, \left[ (i\omega {\bf 1} -{\bf \Gamma})^{-1}\right]_{nn'}
S_{\xi_{m'} \xi_{n'}}.  
        \label{Snn'}\end{eqnarray}
Using Eq.\ (\ref{Sxi}) after simple algebra we obtain
        \begin{eqnarray}
S_{\sigma_m \sigma_n}= 
2  \left[ (-i\omega {\bf 1} -{\bf \Gamma})^{-1}\right]_{mn}\sigma^{st}_n 
        \nonumber \\
+\, 2 \left[ (i\omega {\bf 1} -{\bf \Gamma})^{-1}\right]_{nm}\sigma^{st}_m, 
        \label{Snn'-2}\end{eqnarray}        
which coincides with the result of Fokker-Plank approach (Fourier 
transform of Eq.\ (\ref{K-simple}) without X and Y factors). 

        The technique is similar for the current--current fluctuations.
The case of Eqs.\ (\ref{K-current}) and (\ref{SII}) corresponds to 
currents
        \begin{eqnarray}
X(t)={\tilde e}^j_\pm \left[ \Gamma^j_{m\leftarrow n} \sigma_n (t) 
+\xi^j_{m\leftarrow n} (t) \right] ,
        \nonumber \\
Y(t)={\tilde e}^{j'}_\pm \left[ \Gamma^{j'}_{m'\leftarrow n'} 
\sigma_{n'} (t) +
\xi^{j'}_{m'\leftarrow n'} (t) \right] ,
        \label{XY-cur}\end{eqnarray}
and the straitforward (though rather 
lengthy) calculations using Eqs.\ (\ref{Sxi}) and (\ref{sigma}) lead
to Eq.\ (\ref{SII}). 

        The final expression for the current--occupation spectral 
density is 
        \begin{eqnarray}
S_{X\sigma_k}(\omega )/{\tilde e}^j_\pm = 2 \Gamma^j_{m\leftarrow n}
\left[ (-i\omega {\bf 1}-{\bf \Gamma})^{-1}\right]_{nk} \sigma^{st}_k
        \nonumber \\    
+ 2 \left[ (i\omega {\bf 1}-{\bf \Gamma})^{-1}\right]_{km} 
\Gamma^j_{m\leftarrow n} \sigma^{st}_n \, ,
        \label{SIV}\end{eqnarray}
and it also coincides with the corresponding expression obtained 
in the Fokker-Plank technique.

        Thus, we have proven that the Fokker-Plank method 
is equivalent to the  quasi-Langevin 
 method within the ``orthodox'' framework. However, in
contrast to the former approach, the quasi-Langevin method 
easily allows generalization for the fluctuations in 
the ``quantum'' frequency range. 

Let us remind that in ``orthodox'' theory \cite{Av-Likh}
(from now on for simplicity we speak only about purely 
single-electron tunneling and do not consider cotunneling, 
Andreev reflection, etc.)
        \begin{equation}
\Gamma =\frac{I_0(W/e)}{e(1-\exp (-W/T))}, \,\, W=eV-e^2/2C_{eff}
        \end{equation}
where $I_0(v)$ is the ``seed'' {\it I-V} curve of the junction
(in the linear case $I_0(v)=v/R$), 
$W$ is the energy gain due to tunneling, $V$ is the voltage across
the junction before the tunneling, and $C_{eff}$ is the effective
junction capacitance (which also accounts for the environment).
The generalization of quasi-Langevin method is the substitution of
Eq.\ (\ref{Sxi}) by more general equation \cite{Kor-Bl}
        \begin{eqnarray}
S_{\xi_{m\leftarrow n}^j \xi_{m'\leftarrow n'}^{j'}} (\omega ) = 
  \delta_{mm'}\delta_{nn'}\delta_{jj'} 
[{\tilde \Gamma}^+ +{\tilde \Gamma}^-]  \,
 \sigma_n^{st} \, ,
        \nonumber \\
{\tilde \Gamma}^\pm =
 \frac{I_{0,j}(W^j_{m\leftarrow n} /e \pm \hbar \omega/e)}
{e\left[ 1- \exp \left( ( W^j_{m\leftarrow n} \pm \hbar \omega )/T 
 \right) \right] } .
        \label{Sxi-2}\end{eqnarray}
This equation can be considered as a generalization of the 
fluctuation-dissipation theorem and equations of Ref.\ \cite{Dahm}
for the case of single-electron tunneling. It can be obtained within
the standard tunneling hamiltonian technique averaging the 
quantum current--current
correlator and then taking the Fourier transform.

        Using this generalization one can see that at high frequencies, 
$\omega \gg \Gamma$, the 
occupation--occupation and occupation--current spectral densities
vanish, while for the current--current spectral density Eq.\ 
({\ref{SII}) transforms into 
        \begin{equation}
S_{XY}(\omega ) = {\tilde e}^j_\pm {\tilde e}^{j'}_\pm  
S_{\xi_{m\leftarrow n}^j \xi_{m'\leftarrow n'}^{j'}} (\omega ), 
        \label{SII-2}\end{equation} 
(given by Eq.\ (\ref{Sxi-2})) because the first terms of Eq.\
(\ref{XY-cur}) are too slow to give a cotribution.
This result coincides with the result of Ref.\ \cite{Kor-Bl}.
The advantage of the quasi-Langevin approach is the possibility
to obtain spectral densities in the ``orthodox'' and ``quantum''
frequency ranges using the same formalism while in Ref.\ \cite{Kor-Bl}
they are necessarily treated on different footing. 
        
        In the ``quantum'' frequency range the current
spectral density does not correspond directly to the available power
because of the contribution from zero-point oscillations.
The spectral density of the current calculated above can be 
considered as the power
(within the unit bandwidth) going from the system to a small
external resistance $r$ (divided by $r$ and the coupling factor $\alpha$).
To obtain the available power, we should subtract the power flow in the 
opposite direction which is the product of the voltage spectral
density of the resistance 
$2\hbar\omega r \coth (\hbar \omega/2T_r)$ (the ``receiver'' 
temperature $T_r$ can differ from $T$), the factor $\alpha$,
and the active conductance of the fluctuation source, which in the case
corresponding to Eqs.\ (\ref{Sxi-2})--(\ref{SII-2}) is given by 
\cite{Kor-Bl} 
        \begin{equation}
\mbox{Re} G (\omega) = \left( ({\tilde e}^j_\pm )^2/2\hbar \omega \right) 
[{\tilde \Gamma}^+ -
{\tilde \Gamma}^-] 
        \label{conductance}\end{equation} 
(this expression obviously corresponds to the lowest order of 
photon-assisted processes).
Hence, the contribution to be subtracted from $S_{\xi_{m\leftarrow n}^j}$
is equal to $2\hbar\omega \coth (\hbar \omega/2T_r) \mbox{Re} 
G(\omega )$. Traditionally this contribution is called zero-point for 
$T_r=0$. However, for finite temperature $T$ a more natural choice
of the ``receiver'' temperature is $T_r=T$. Notice that to get the 
total conductance Re$G_t(\omega)$, Eq.\ (\ref{conductance}) should 
be summed over all kinds of tunneling events. 

        The solid line 1 in Fig.\ \ref{S(w)} shows the numerical 
result for the spectral 
density $S_{II}(\omega )$ of the current in the external lead of the 
single-electron transistor consisting of two similar tunnel junctions
with capacitances $C_1=C_2=C_\Sigma /2$ and resistances 
$R_1=R_2=R_\Sigma /2$ at $Q_0=0.3e$ and $T=0.03e^2/C_\Sigma$. 
        The frequency dependence in the ``orthodox'' frequency range
is important at $f=\omega /2\pi \sim I/e$ while the ``quantum'' frequency
dependence occurs at $f \sim e^2/C_\Sigma h$. Because the junction 
resistances are chosen sufficiently large, $R_\Sigma=100R_K$, two
frequency ranges are far away from each other (the ratio of typical
frequencies is on the order of $R/R_K$).
        The line 2 shows the spectral density corresponding
to the power available at zero-temperature receiver (zero-point 
contribution is subtracted, $T_r=0$) while the line 3 shows the 
available power for $T_r=T$.
	Notice that the curve 3 differs considerably from curve 1
even in the ``orthodox'' frequency range because of thermodynamical
reasons. 
	The dashed lines 4 and 5 show the results for $T=0$ (two
lower curves obviously coincide). Notice the cusps at 
$hf=0.05e^2/C_\Sigma$ which correspond to the extra energy above
the Coulomb blockade and the abrupt vanishing of the available
power above the frequency $f=W_{max}/h$ ($f=0.45e^2/C_\Sigma h$ in 
this case) where $W_{max}$ is the maximal energy gain among all 
possible tunneling events.
    
        In the case $R_j \gg R_K$  the ``orthodox''
and ``quantum'' frequency ranges are
far apart from each other because $\Gamma \sim W/eR \ll W/h$.
The fact that the quasi-Langevin formalism describes the fluctuations
for the whole frequency axis, suggests the possibility to use this
method also when two frequency ranges are close to each other. 
However, this is impossible because when $R_j \sim R_K$ the master
equation approach fails due to strong cotunneling processes 
\cite{Av-Odin}. The correct description of the noise in this case
(as well as just calculation of the average currents) is still 
an open question despite an important progress for the average
currents \cite{Schon}.

        In conclusion, we developed the quasi-Langevin method
for the calculation of the fluctuations in single-electron tunneling.
Its equivalence to the previous Fokker-Plank approach in the
``orthodox'' frequency range is proven. The advantage of the 
quasi-Langevin method is the natural generalization for the 
``quantum'' frequency range.

        The author thanks K.\ K.\ Likharev for attracting the interest
to this problem and N.\ E.\ Korotkov for the basic introduction to
the theory of fluctuations. The author is also grateful to D.\ V.\
Averin, S.\ V.\ Gantsevich, Sh.\ M.\ Kogan, V.\ V.\ Potemkin, G.\ Sch\"{o}n,  
A.\ S.\ Stepanov, and J.\ S.\ Tsai for the valuable discussions.
        The work was supported in parts by US AFOSR, Russian Fund for 
Basic Research, and Russian Program on Prospective Technologies for 
Nanoelectronics.

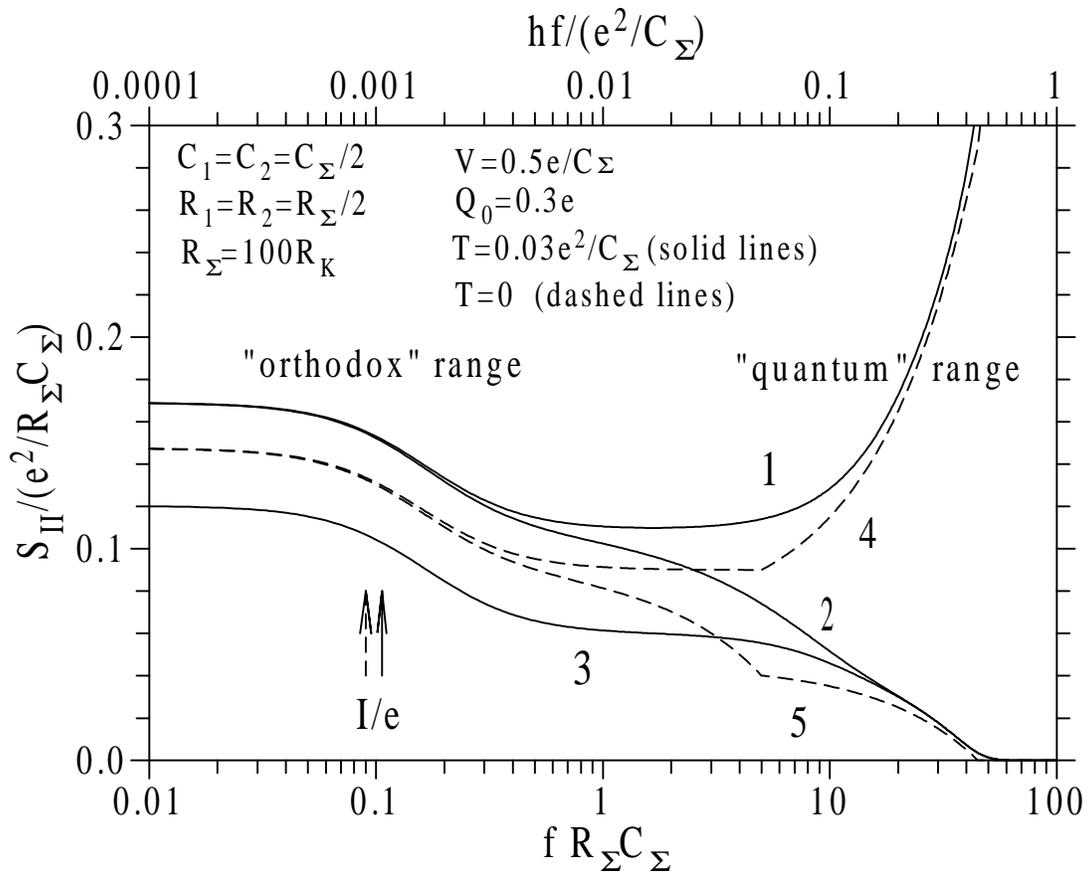
        \begin{figure}
\caption{ Curve 1 shows the spectral density of the current 
(in the external lead) through the symmetric single-electron 
transistor at $T=0.03e^2/C_\Sigma$ as a
function of frequency $f=\omega /2\pi$. The relatively large 
junction resistances provide cosiderable difference between 
``orthodox'' and ``quantum'' frequency scales. The curve 2
shows the spectral density corresponding
to the available power with zero-point cotribution subtracted 
($T_r=0$) while for the curve 3 the ``receiver'' is assumed to be 
at the same temperature ($T_r=T$). Dashed curves 4 and 5 
demonstrate the results for $T=0$. The average current 
$I=0.106e/R_\Sigma C_\Sigma$ for $T=0.03e^2/C_\Sigma$ while
$I=0.09e/R_\Sigma C_\Sigma$ for $T=0$. }
\label{S(w)}\end{figure}


\begin{references}
\bibitem{complicat} Notice that in several later papers instead 
        of this simple method, 
        a more complicated method which requires the
        knowledge of all eigenstates and eigenfunctions of the
        evolution matrix have been used.
\end{references}
\end{document}